\documentstyle[12pt]{article}
\newcommand{\nc}{\newcommand}
\nc{\renc}{\renewcommand}

\nc{\etal}{\mbox{\it et al. }}
\nc{\ie}{{\it i.e.}}
\nc{\eg}{{\it e.g.}}

\renc{\thefootnote}{\arabic{footnote}}
\nc{\capt}[1]{{\bf Figure.} {\small\sl #1}}

\nc{\eqs}[2]{\mbox{Eqs.~(\ref{#1},\,\ref{#2})}}
\nc{\eq}[1]{\mbox{Eq.~(\ref{#1})}}

\nc{\figs}[2]{\mbox{Figs.~(\ref{#1},\,\ref{#2})}}
\nc{\fig}[1]{\mbox{Fig~.(\ref{#1})}}

\nc{\tag}[1]{\label{#1} \marginpar{{\footnotesize #1}}}
\nc{\mtag}[1]{\label{#1} \mbox{\marginpar{{\footnotesize #1}}}}
\renc{\baselinestretch}{1.2}
\newlength{\overeqskip}
\newlength{\undereqskip}
\nc{\be}[1]{\begin{equation} \mbox{$\label{#1}$}}
\nc{\bea}[1]{\begin{eqnarray} \mbox{$\label{#1}$}}
\nc{\Section}[2]{\section{#2}\label{#1}}
\nc{\Bibitem}[1]{\bibitem{#1}}
\nc{\Label}[1]{\label{#1}}
\nc{\eea}{\vspace{\undereqskip}\end{eqnarray}}
\nc{\ee}{\vspace{\undereqskip}\end{equation}}
\nc{\bdm}{\begin{displaymath}}
\nc{\edm}{\end{displaymath}}
\nc{\dpsty}{\displaystyle}
\nc{\bc}{\begin{center}}
\nc{\ec}{\end{center}}
\nc{\ba}{\begin{array}}
\nc{\ea}{\end{array}}
\nc{\bab}{\begin{abstract}}
\nc{\eab}{\end{abstract}}
\nc{\btab}{\begin{tabular}}
\nc{\etab}{\end{tabular}}
\nc{\bit}{\begin{itemize}}
\nc{\eit}{\end{itemize}}
\nc{\ben}{\begin{enumerate}}
\nc{\een}{\end{enumerate}}
\nc{\bfig}{\begin{figure}}
\nc{\efig}{\end{figure}}
\nc{\arreq}{&\!=\!&}
\nc{\arrmi}{&\!-\!&}
\nc{\arrpl}{&\!+\!&}
\nc{\arrap}{&\!\!\!\approx\!\!\!&}
\nc{\non}{\nonumber\\*}
\nc{\align}{\!\!\!\!\!\!\!\!&&}

\def\lsim{\; \raise0.3ex\hbox{$<$\kern-0.75em
      \raise-1.1ex\hbox{$\sim$}}\; }
\def\gsim{\; \raise0.3ex\hbox{$>$\kern-0.75em
      \raise-1.1ex\hbox{$\sim$}}\; }
\nc{\DOT}{\hspace{-0.08in}{\bf .}\hspace{0.1in}}
\nc{\Laada}{\hbox {$\sqcap$ \kern -1em $\sqcup$}}
\nc\loota{{\scriptstyle\sqcap\kern-0.55em\hbox{$\scriptstyle\sqcup$}}}
\nc\Loota{{\sqcap\kern-0.65em\hbox{$\sqcup$}}}
\nc\laada{\Loota}
\nc{\qed}{\hskip 3em \hbox{\BOX} \vskip 2ex}
\def\Re{{\rm Re}\hskip2pt}

\nc{\real}{{\rm I \! R}}
\nc{\Z}{{\sf Z \!\!\! Z}}
\nc{\complex}{{\rm C\!\!\! {\sf I}\,\,}}
\def\bigid{\leavevmode\hbox{\small1\kern-3.8pt\normalsize1}}
\def\id{\leavevmode\hbox{\small1\kern-3.3pt\normalsize1}}
\nc{\slask}{\!\!\!/}
\nc{\bis}{{\prime\prime}}
\nc{\pa}{\partial}
\nc{\na}{\nabla}
\nc{\ra}{\rangle}
\nc{\la}{\langle}
\nc{\goto}{\rightarrow}
\nc{\swap}{\leftrightarrow}

\nc{\EE}[1]{ \mbox{$\cdot10^{#1}$} }
\nc{\abs}[1]{\left|#1\right|}
\nc{\at}[2]{\left.#1\right|_{#2}}
\nc{\norm}[1]{\|#1\|}
\nc{\abscut}[2]{\Abs{#1}_{\scriptscriptstyle#2}}
\nc{\vek}[1]{{\rm\bf #1}}
\nc{\integral}[2]{\int\limits_{#1}^{#2}}
\nc{\inv}[1]{\frac{1}{#1}}
\nc{\dd}[2]{{{\partial #1}\over{\partial #2}}}
\nc{\ddd}[2]{{{{\partial}^2 #1}\over{\partial {#2}^2}}}
\nc{\dddd}[3]{{{{\partial}^2 #1}\over
	{\partial #2 \partial #3}}}
\nc{\dder}[2]{{{d #1}\over{d #2}}}
\nc{\ddder}[2]{{{d^2 #1}\over{d {#2}^2}}}
\nc{\dddder}[3]{{d^2 #1}\over
	{d #2 d #3}}
\nc{\dx}[1]{d\,^{#1}x}
\nc{\dy}[1]{d\,^{#1}y}
\nc{\dz}[1]{d\,^{#1}z}
\nc{\dl}[1]{\frac{d\,^{#1}l}{(2\pi)^{#1}}}
\nc{\dk}[1]{\frac{d\,^{#1}k}{(2\pi)^{#1}}}
\nc{\dq}[1]{\frac{d\,^{#1}q}{(2\pi)^{#1}}}

\nc{\cc}{\mbox{$c.c.$ }}
\nc{\hc}{\mbox{$h.c.$ }}
\nc{\cf}{cf.\ }
\nc{\erfc}{{\rm erfc}}
\nc{\Tr}{{\rm Tr\,}}
\nc{\tr}{{\rm tr\,}}
\nc{\pol}{{\rm pol}}
\nc{\sign}{{\rm sign}}
\nc{\bfT}{{\bf T }}

\nc{\cA}{{\cal A}}
\nc{\cB}{{\cal B}}
\nc{\cD}{{\cal D}}
\nc{\cE}{{\cal E}}
\nc{\cG}{{\cal G}}
\nc{\cH}{{\cal H}}
\nc{\cL}{{\cal L}}
\nc{\cO}{{\cal O}}
\nc{\cT}{{\cal T}}
\nc{\cN}{{\cal N}}
\nc{\cI}{{\cal I}}
\nc{\rvac}[1]{|{\cal O}#1\rangle}
\nc{\lvac}[1]{\langle{\cal O}#1|}
\nc{\rvacb}[1]{|{\cal O}_\beta #1\rangle}
\nc{\lvacb}[1]{\langle{\cal O}_\beta #1 |}
\nc{\bb}{\bar{\beta}}
\nc{\bt}{\tilde{\beta}}
\nc{\ctH}{\tilde{\cal H}}
\nc{\chH}{\hat{\cal H}}
\nc{\1}{\aa}
\nc{\2}{\"{a}}
\nc{\3}{\"{o}}
\nc{\4}{\AA}
\nc{\5}{\"{A}}
\nc{\6}{\"{O}}
\nc{\al}{\alpha}
\nc{\g}{\gamma}
\nc{\Del}{\Delta}
\nc{\e}{\epsilon}
\nc{\eps}{\epsilon}
\nc{\lam}{\lambda}
\nc{\om}{\omega}
\nc{\Om}{\Omega}
\nc{\ve}{\varepsilon}
\nc{\mn}{{\mu\nu}}
\nc{\k}{\kappa}
\nc{\vp}{\varphi}

\nc{\advp}[3]{{\it  Adv.\ in\ Phys.\ }{{\bf #1} {(#2)} {#3}}}
\nc{\annp}[3]{{\it  Ann.\ Phys.\ (N.Y.)\ }{{\bf #1} {(#2)} {#3}}}
\nc{\apl}[3]{{\it  Appl. Phys. Lett. }{{\bf #1} {(#2)} {#3}}}
\nc{\apj}[3]{{\it  Ap.\ J.\ }{{\bf #1} {(#2)} {#3}}}
\nc{\apjl}[3]{{\it  Ap.\ J.\ Lett.\ }{{\bf #1} {(#2)} {#3}}}
\nc{\app}[3]{{\it Astropart.\ Phys.\ }{{\bf #1} {(#2)} {#3}}}  
\nc{\cmp}[3]{{\it  Comm.\ Math.\ Phys.\ }{{ \bf #1} {(#2)} {#3}}}
\nc{\cqg}[3]{{\it  Class.\ Quant.\ Grav.\ }{{\bf #1} {(#2)} {#3}}}
\nc{\epl}[3]{{\it  Europhys.\ Lett.\ }{{\bf #1} {(#2)} {#3}}}
\nc{\ijmp}[3]{{\it Int.\ J.\ Mod.\ Phys.\ }{{\bf #1} {(#2)} {#3}}}
\nc{\ijtp}[3]{{\it Int.\ J.\ Theor.\ Phys.\ }{{\bf #1} {(#2)} {#3}}}
\nc{\jmp}[3]{{\it  J.\ Math.\ Phys.\ }{{ \bf #1} {(#2)} {#3}}}
\nc{\jpa}[3]{{\it  J.\ Phys.\ A\ }{{\bf #1} {(#2)} {#3}}}
\nc{\jpc}[3]{{\it  J.\ Phys.\ C\ }{{\bf #1} {(#2)} {#3}}}
\nc{\jap}[3]{{\it J.\ Appl.\ Phys.\ }{{\bf #1} {(#2)} {#3}}}
\nc{\jpsj}[3]{{\it J.\ Phys.\ Soc.\ Japan\ }{{\bf #1} {(#2)} {#3}}}
\nc{\lmp}[3]{{\it Lett.\ Math.\ Phys.\ }{{\bf #1} {(#2)} {#3}}}
\nc{\mpl}[3]{{\it  Mod.\ Phys.\ Lett.\ }{{\bf #1} {(#2)} {#3}}}
\nc{\ncim}[3]{{\it  Nuov.\ Cim.\ }{{\bf #1} {(#2)} {#3}}}
\nc{\np}[3]{{\it  Nucl.\ Phys.\ }{{\bf #1} {(#2)} {#3}}}
\nc{\pr}[3]{{\it Phys.\ Rev.\ }{{\bf #1} {(#2)} {#3}}}
\nc{\pra}[3]{{\it  Phys.\ Rev.\ A\ }{{\bf #1} {(#2)} {#3}}}
\nc{\prb}[3]{{\it  Phys.\ Rev.\ B\ }{{{\bf #1} {(#2)} {#3}}}}
\nc{\prc}[3]{{\it  Phys.\ Rev.\ C\ }{{\bf #1} {(#2)} {#3}}}
\nc{\prd}[3]{{\it  Phys.\ Rev.\ D\ }{{\bf #1} {(#2)} {#3}}}
\nc{\prl}[3]{{\it Phys\ Rev.\ Lett.\ }{{\bf #1} {(#2)} {#3}}}
\nc{\pl}[3]{{\it  Phys.\ Lett.\ }{{\bf #1} {(#2)} {#3}}}
\nc{\prep}[3]{{\it Phys\. Rep.\ }{{\bf #1} {(#2)} {#3}}}
\nc{\prsl}[3]{{\it Proc.\ R.\ Soc.\ London\ }{{\bf #1} {(#2)} {#3}}}
\nc{\ptp}[3]{{\it  Prog.\ Theor.\ Phys.\ }{{\bf #1} {(#2)} {#3}}}
\nc{\ptps}[3]{{\it  Prog\ Theor.\ Phys.\ suppl.\ }{{\bf #1} {(#2)} {#3}}}
\nc{\physa}[3]{{\it  Physica\ A\ }{{\bf #1} {(#2)} {#3}}}
\nc{\physb}[3]{{\it  Physica\ B\ }{{\bf #1} {(#2)} {#3}}}
\nc{\phys}[3]{{\it Physica\ }{{\bf #1} {(#2)} {#3}}}
\nc{\rmp}[3]{{\it  Rev.\ Mod.\ Phys.\ }{{\bf #1} {(#2)} {#3}}}
\nc{\rpp}[3]{{\it Rep.\ Prog.\ Phys.\ }{{\bf #1} {(#2)} {#3}}}
\nc{\sjnp}[3]{{\it Sov.\ J.\ Nucl.\ Phys.\ }{{\bf #1} {(#2)} {#3}}}
\nc{\spjetp}[3]{{\it Sov.\ Phys.\ JETP\ }{{\bf #1} {(#2)} {#3}}}
\nc{\yf}[3]{{\it Yad.\ Fiz.\ }{{\bf #1} {(#2)} {#3}}}
\nc{\zetp}[3]{{\it Zh.\ Eksp.\ Teor.\ Fiz.\  }{{\bf #1}  {(#2)} {#3}}}
\nc{\zp}[3]{{\it Z.\ Phys.\ }{{\bf #1} {(#2)} {#3}}}
\nc{\ibid}[3]{{\sl ibid.\ }{{\bf #1} {#2} {#3}}}
\nc{\rf}[1]{(\ref{#1})}
\nc{\nn}{\nonumber \\*}
\nc{\bfB}{\bf{B}}
\nc{\bfv}{\bf{v}}
\nc{\bfx}{\bf{x}}
\nc{\bfy}{\bf{y}}
\nc{\vx}{\vec{x}}
\nc{\vy}{\vec{y}}
\nc{\oB}{\overline{B}}
\nc{\oI}{\overline{I}}
\nc{\oR}{\overline{R}}
\nc{\rar}{\rightarrow}
\nc{\ti}{\times}
\nc{\slsh}{\hskip-5pt/}
\nc{\sm}{Standard~Model~}
\nc{\MP}{M_{\rm Pl}}
\nc{\tp}{t_{\rm Pl}}
\nc{\ave}{\bar{E}}

\renc{\min}{p_{\rm min}}
\renc{\max}{p_{\rm max}}
\nc{\pmin}{p_{\rm min}}
\nc{\pmax}{p_{\rm max}}
\nc{\fo}{f_0}
\nc{\foi}{f_{0,i}\,}
\nc{\fop}{f_0^P}
\nc{\fou}{f_0^U}
\def\sepand{\rule{14cm}{0pt}\and}
\nc{\eff}{{\rm eff}}
\nc{\MT}{M_{\rm T}}
\nc{\ML}{M_{\rm L}}
\nc{\kk}{\vek{k}}
\nc{\pp}{{\rm p}}
\nc{\cb}{critical bubble~}
\nc{\cbs}{critical bubbles~}
\nc{\scb}{subcritical bubble~}
\nc{\scbs}{subcritical bubbles~}
\nc{\MSSM}{Minimal Supersymmetric Standard Model}
\nc{\mato}[1]{\mathaccent 126 #1}
\nc{\viiva}[1]{\overline{#1}}
\nc{\D}{\Delta}
\nc{\arccot}{{\rm arccot}}

\begin{document} 

{\title{{\hfill {{\small  TURKU-FL-P31-98}}\vskip 1truecm}
{\bf Electroweak Baryogenesis in a Left-Right Supersymmetric Model}}

\vspace{1.2cm}
 
\author{
{\sc Tuomas Multam\" aki$^{1}$}\\
{\sl and}\\
{\sc Iiro Vilja$^{2}$ }\\ 
{\sl Department of Physics,
University of Turku} \\
{\sl FIN-20014 Turku, Finland} \\
\sepand
}
\date{October 9, 1998}
\maketitle}
\vspace{2cm}
%\newpage
\begin{abstract}
\noindent 
The possibility of electroweak baryogenesis is considered within
the framework of a left-right supersymmetric model. It is shown
that for a range of parameters the large
sneutrino VEV required for parity breaking varies
at the electroweak phase transition leading to a production
of baryons. The resulting baryon to entropy ratio is approximated to be
${n_B\over s}\sim \alpha\ 0.7 \times10^{-8}$, where $\alpha$ is the 
angle that the phase of sneutrino VEV changes at the electroweak phase
transition.
\end{abstract}
\vfill
\footnoterule
{\small$^1$tuomul@newton.tfy.utu.fi,  $^2$vilja@newton.tfy.utu.fi}

\thispagestyle{empty}
\newpage
\setcounter{page}{1}
%%%%%%%%%%%%%%%%%%%%%%%%%%%%%%%%%%%%%%%%%%%%%%%%%%%%%%%%%%%%%
The Standard Model (SM) fulfills all the requirements,
including the Sakharov conditions \cite{sakh}, for
electroweak baryogenesis \cite{rev} that may be responsible
for the observed baryon asymmetry of the universe.
However, 
in the SM the phase transition has been shown to be too weakly
first order to preserve the generated asymmetry \cite{shapo}.
Furthermore, the phase of the Cobayashi-Maskawa -matrix leads
to an insufficient CP-violation \cite{gavela} which further
indicates that physics beyond the SM is needed to account for
the baryon asymmetry.

One of the most popular extensions of the SM is the Minimal Supersymmetric
Standard Model (MSSM) that, with an appropriate choice of parameters,
leads to a large enough baryon asymmetry, ${n_B\over s}\sim10^{-10}$
\cite{strength1, CQRVW}. Some of the parameter bounds are, however,
quite stringent \cite{multa1} (although Riotto has recently argued
\cite{riotto2} that an additional enchancement factor of about $10^2$ is
present). 

In the MSSM the conservation of R-parity, $R=(-1)^{3(B-L)+2S}$, is
assumed and put in by hand. This is quite {\em ad hoc} since it is
not required for the internal consistency of the model but 
for the conservation of baryon and lepton numbers. 
To avoid {\it e.g.} proton decay, the R-parity breaking
terms, allowed by gauge invariance, are therefore usually omitted in the MSSM 
or their couplings are assumed to be very small.

The supersymmetric left-right model (SUSYLR) based on the gauge group
$SU(2)_L\times SU(2)_R\times U(1)_{B-L}$ explicitly conserves 
R-parity \cite{rparit}. Considering the
MSSM to be a low energy approximation
of the SUSYLR it can also account for the baryon asymmetry of the
universe. 
Furthermore, the see-saw mechanism \cite{seesaw} that can be 
accomodated in the left-right models, can be used to generate
light masses for left-handed neutrinos and large masses for right-handed
ones. The possibility of a small (left-handed) neutrino mass   
is therefore often seen as another motivation for the SUSYLR model. 

By viewing the MSSM as a low energy approximation of the SUSYLR it
becomes warranted to question whether some properties of the SUSYLR
affect the prospect of electroweak baryogenesis. This question is studied
in the context of the present paper and it is argued that electroweak baryogenesis
may be significantly larger in the SUSYLR than in the MSSM. 

The gauge-invariant superpotential for the SUSYLR model with two
bidoublets \cite{superp} at zero temperature
is given by (omitting all generation and $SU(2)$ indices and the Q-terms):
\bea{superpot}
W & = & h_\phi L^T\tau_2\Phi\tau_2L^c+h_\chi L^T\tau_2\chi\tau_2L^c\nonumber\\
& + & if(L^T\tau_2\Delta L+L^{cT}\tau_2\Delta^cL^c)\nonumber\\
& + & M\ \tr(\Delta\viiva{\Delta}+\Delta^c\viiva{\Delta^c})+
\mu_1\ \tr(\tau_2\Phi^T\tau_2\chi)\nonumber\\
& + & \mu'_1\ \tr(\tau_2\Phi^T\tau_2\Phi)+\mu''_1\ \tr(\tau_2\chi^T\tau_2\chi),
\eea
where $L(2,1,-1),\ L^c(1,2,1)$ are the matter 
superfields and $\Delta(3,1,2)$, $\Delta^c(1,3,-2)$, 
$\viiva{\Delta}(3,1,-2)$, $\viiva{\Delta^c}(1,3,2)$, 
$\Phi(2,2,0)$, $\chi(2,2,0)$ are the Higgs 
superfields with their respective $SU(2)_L\times SU(2)_R\times U(1)_{B-L}$
quantum numbers. 
It is worth noting that since one of the bidoublets will be responsible
for the electron mass the mechanism considered here requires at least
two bidoublet fields to result in a significant production of baryons.

The scalar components of the superfields have the following 
component representations:
\be{charges}
\mato{L}={\mato{\nu}\choose \mato{e}^-},\ 
\mato{L^c}={\mato{\nu^c}\choose \mato{e}^+},\
\Delta=\left(\matrix{
{\delta^+\over\sqrt{2}} & \delta^{++} \cr
\delta^0                & -{\delta^+\over\sqrt{2}} \\
}\right), 
\Phi=\left(\matrix{
\phi_1^0 & \phi_2^+\cr
\phi_1^- & \phi_2^0 \\
}\right),
\chi=\left(\matrix{
\chi_1^0 & \chi_2^+\cr
\chi_1^- & \chi_2^0 \\
}\right),
\ee
and similarly for the rest of the fields.

By standard methods \cite{haberkane} it can be shown that the 
Lagrangean contains the following terms (omitting generation indices):
\be{kytke}
h_\chi \viiva{\mato{H}^c}\mato{\nu}_R^*L e,
\ee
where $H={\chi_2^+\choose \chi_1^-}$ and L is the left-handed chirality
projection operator. Of cource similar term proportional to $h_\phi$
is present in the Lagrangean but since $h_\phi$ is proportional to 
the electron mass it is neglected here \cite{huitu}.
The Yukawa coupling $h_\chi$, however, is proportional to the neutrino Dirac mass
and is hence only bounded by an experimental upper limit \cite{huitu}.

Comparing with ref. \cite{multa2} we note that 
this is exactly of the form that creates a CP-violating source
in the diffusion equation at the electroweak (EW) phase transition assuming
that $<\mato{\nu}>$ does not vanish and is not a constant. 
It has been shown that in a wide class of SUSYLR models R-parity
is necessarily spontaneously broken by a large sneutrino VEV \cite{kuchi}.
Thus for electroweak baryogenesis we need to show that $<\mato{\nu}>$ 
varies at the EW phase transition. Following ref. 
\cite{multa2} the CP-violating source can now be computed
by using methods described in \cite{riotto}. 

In electroweak baryogenesis the created
baryon to entropy ratio is 
\be{nbovers}
{n_B\over s}=-g(k_i){\cA\viiva{D}\Gamma_{ws}\over v_w^2s},
\ee
where $g(k_i)$ is a numerical coefficient depending on the degrees of 
freedom, $\viiva{D}$ the effective diffusion rate, $\Gamma_{ws}=6
\alpha_w^4 T (\kappa=1)$ the weak sphaleron rate and $v_w$ the 
speed of the bubble wall. In \cite{multa2} it was shown that 
\bea{int1}
\cA\sim I_1 & \equiv & 2 h_{ij}h_{kj}\int_0^\infty du\Big[\sin(\theta_i-
\theta_k)[A_iA'_k-A_kA_i']\nonumber \\
& - & \cos(\theta_i-\theta_k)A_iA_k(\theta_i+\theta_k)'\Big]
e^{-\lambda_+u}\cI^{e_j}_{\mato{H}},
\eea
where $\lambda_+=(v_w+\sqrt{v_w^2+4\mato{\Gamma}\viiva{D}})/(2\viiva{D})$
and $<\mato{\nu}>\equiv Ae^{i\theta}$.
Clearly ${n_B\over s}$ vanishes unless $A$ and $\theta$ change when
going over the bubble wall.

To show that electroweak baryogenesis is indeed possible in this model 
we must 
first consider the associated Higgs potential.
The Higgs potential including soft breaking terms
(omitting the $Q$ - terms again) is \cite{superp}: 
\be{higpot}
V = V_{F-terms} ~~+ V_{soft} ~~+ V_{D-terms}
\ee
where
\bea{hig1}
V_{F-terms} & = & \mid \tilde{L}^{T}
\tau_{2} (h_\phi\Phi+h_\chi\chi) \tau_{2} + 2 i f \tilde{L}^{cT} \tau_{2} \Delta^{c}\mid^{2}\nonumber\\
& + & \mid \tilde{L}^{cT} \tau_{2} (h_\phi\Phi+h_\chi\chi)^T
\tau_{2} + 2 i f \tilde{L}^{T} \tau_{2} \D\mid^{2}
+ \tr \mid h_\phi \tilde{L}^{c} \tilde{L}^{T} + 2 \mu_1' \Phi^{T}+\mu_1\chi^T\mid^{2}\nonumber\\
& + & \tr \mid h_\chi \tilde{L}^{c} \tilde{L}^{T} + 2 \mu_1'' \chi^{T}+\mu_1\Phi^T\mid^{2}
+ \tr \mid i f \tilde{L} \tilde{L}^{T} \tau_{2} + M \overline{\D}\mid^{2}
\nonumber \\
& + & \tr \mid i f \tilde{L}^{c}
\tilde{L}^{cT} \tau_{2} + M \overline{\D}^{c}\mid^{2}
+ \mid M \mid^{2} \tr \left ( \Delta^{\dagger} \Delta+ \Delta^{c\dagger}\Delta^{c}\right),
\eea
\bea{hig2}
V_{soft} & = & m_{l}^{2} \left (
\tilde{L}^{\dagger} \tilde{L} + \tilde{L}^{c\dagger} \tilde{L}^{c} \right )
+ \left (M_{1}^{2} - \mid M \mid^{2} \right ) \tr \left (\Delta^{\dagger} \D
+ \Delta^{c\dagger} \Delta^{c} \right )\nonumber \\
& + & \left (M_{2}^{2} - \mid M \mid^{2} \right )
\tr \left (\overline{\D}^{\dagger} \overline{\D}
+ \overline{\D}^{c\dagger} \overline{\D}^{c} \right )
+ \left (M'^{2} \tr \left ( \Delta\overline{\D} + \Delta^{c} \overline{\D}^{c}
\right ) + h. c. \right ) \nonumber \\
& + & \left ( M_{\phi}^{2} - 4 \mid \mu_1' \mid^{2} \right )
\tr \Phi^{\dagger} \Phi+
\left ( M_{\chi}^{2} - 4 \mid \mu_1'' \mid^{2} \right )
\tr \chi^{\dagger} \chi)\nonumber\\
& + & [\left ( M_{\phi\chi}^{2} - 4 \mid \mu_1 \mid^{2} \right 
)\tr \Phi^{\dagger}\chi+\hc]\nonumber\\
& + & \left ( {{\mu_2^{2}} \over 2} \tr \left (\tau_{2} \Phi^{T} \tau_{2} \chi
\right )+
{{\mu_2'^{2}} \over 2} \tr \left (\tau_{2} \Phi^{T} \tau_{2} \Phi
\right ) +
{{\mu_2''^{2}} \over 2} \tr \left (\tau_{2} \chi^{T} \tau_{2} \chi
\right ) + h. c. \right ) \nonumber \\
& + & \left ( i v \left (\tilde{L}^{T}
\tau_{2} \Delta\tilde{L} + \tilde{L}^{cT} \tau_{2} \Delta^{c} \tilde{L}^{c}
\right ) + \tilde{L}^{T}
\tau_{2} (e_1\Phi+e_2\chi) \tau_{2} \tilde{L}^{c} + h.c. \right ),
\eea
\bea{hig3}
V_{D-terms} & = & {{g_L^{2}} \over 8}
\sum_{m} \mid \tilde{L}^{\dagger} \tau_{m} \tilde{L}
+ \tr \left ( 2 \Delta^{\dagger} \tau_{m} \Delta+ 2 \overline{\D}^{\dagger} \tau_{m}
\overline{\D} + \Phi^{\dagger} \tau_{m} \Phi +\chi^{\dagger} \tau_{m} \chi\right ) \mid^{2} \nonumber \\
& + & {{g_R^{2}} \over 8}
\sum_{m} \mid \tilde{L}^{c\dagger} \tau_{m} \tilde{L}^{c}
+ \tr \left ( 2 \Delta^{c\dagger} \tau_{m} \Delta^{c}
+ 2 \overline{\D}^{c\dagger} \tau_{m} \overline{\D}^{c}
+ \Phi \tau_{m}^{T} \Phi^{\dagger}+ \chi \tau_{m}^{T} \chi^{\dagger} \right ) \mid^{2} \nonumber \\
& + & {{g'^{2}} \over 8} \mid
\tilde{L}^{c\dagger} \tilde{L}^{c} - \tilde{L}^{\dagger} \tilde{L} +
2 \tr \left ( \Delta^{\dagger} \Delta- \Delta^{c\dagger} \Delta^{c} - \overline{\D}^{\dagger}
\overline{\D} + \overline{\D}^{c\dagger}\overline{\D}^{c}\right ) \mid^{2}.
\eea
In \cite{kuchi}
it was shown that the spontaneous breakdown of parity cannot occur 
without spontaneous R-parity breaking in a model with one bidoublet. 
We shall now examine the Higgs 
potential using assumptions similar to those in \cite{kuchi}. 
We assume that $g^2$ and $g'^2$ are much smaller than $h^2$ and $f^2$ and 
to ensure the required hierarchy of parity and R-parity 
breaking scales, the couplings involving 
$\Phi$ or $\chi$ are assumed to be much smaller than those not involving 
them. With these approximations the potential simplifies to the same
form as in \cite{kuchi}:
\bea{snupot}
V & = & m_{l}^{2} \left (
\tilde{L}^{\dagger} \tilde{L} + \tilde{L}^{c\dagger} \tilde{L}^{c} \right )
+ M_{1}^{2} ~Tr \left ( \D^{\dagger} \D + \D^{c\dagger} \D^{c} \right )
+ M_{2}^{2} ~Tr \left ( \overline{\D}^{\dagger} \overline{\D}
+ \overline{\D}^{c\dagger} \overline{\D}^{c} \right ) \nonumber \\
& + & \mid h \mid^{2} \tilde{L}^{c\dagger}
\tilde{L}^{c} \tilde{L}^{\dagger} \tilde{L} + \mid f \mid^{2}
\left ( \left ( \tilde{L}^{\dagger} \tilde{L} \right )^{2} +
\left (\tilde{L}^{c\dagger} \tilde{L}^{c}\right )
^{2} \right )\nonumber \\
& + & 4 \mid f \mid^{2} \left ( \mid \tilde{L}^{cT} \tau_{2} \D^{c} \mid^{2}
+ \mid \tilde{L}^{T} \tau_{2} \D \mid^{2}\right ) + M'^{2}~Tr\left (
\D \overline {\D} + \D^{c} \overline{\D}^{c} + h.c. \right )
\nonumber \\
& + & \left (\tilde{L}^{T} \tau_{2} \left (
i v \D + i M^{*} f \overline{\D}^{\dagger} \right ) \tilde{L}
+ \tilde{L}^{cT} \tau_{2} \left ( i v \D^{c} +
i M^{*} f \overline{\D}^{c\dagger} \right ) \tilde{L}^{c} + h.c. \right ).
\eea
We will now select the VEVs of the slepton fields to be of the form
\be{vevs}
\left<\tilde{L}^{c}\right> = \left ( \begin{array}{c}
             l' \\
             0
         \end{array} \right ),\ \left<\tilde{L}\right> = \left ( \begin{array}{c}
             l \\
             0
         \end{array} \right )
\ee
and using $SU(2)_L\times SU(2)_R$-invariance these can be chosen real.
In \cite{kuchi} it was shown that for the absolute minimum of the potential, the
triplet fields must be of the $Q_{em}$ preserving form 
\bea{triplets}
\left<\D^{c}\right> = 
		\left(
		\begin{array}{cc}
                0 & 0\\
                \delta' & 0
        	\end{array}
                \right) & , & 
\left<\D\right> =
		\left(
		\begin{array}{cc}
                0 & 0\\
                \delta & 0
                \end{array}
                \right )\nonumber\\
\left<\overline{\D}^{c}\right>= 
		\left(\begin{array}{cc}
                0 & \overline{\delta'}\\
                0 & 0
                \end{array}
                \right) & , & 
\left<\overline{\D}\right>=
		\left(\begin{array}{cc}
                0 & \overline{\delta}\\
                0 & 0
                \end{array}
                \right ).
\eea
The doublet and triplet VEVs can now be determined by substituting 
(\ref{vevs}) and (\ref{triplets}) into (\ref{snupot}) and finding the
minimum. The algebra is quite involved and the general solutions are
messy so that it is useful to approximate $M_1^2$ and $M'^2$ as
small perturbations as was done in \cite{kuchi}.
The potential now takes the form
\bea{pot2}
V & = & m_l^2(l^2+l'^2)+m_2^2(l^2+l'^2)+
\abs{h}^2 l^2\nonumber\\
& + & \abs{f}^2(l^4+l^4)\nonumber\\
& + & [l^2(v\delta+fM^*\viiva{\delta}^*)+l'^2(v\delta'+fM^*\viiva{\delta}'^*)+c.c].
\eea
The parity breaking solution
is \cite{kuchi}
\bea{sol}
l & = & \overline{\delta} =  0,\ \delta' = - {v \over {4 
f^{2}}},\ 
\overline{\delta}' = - {{f M l'^{2}} \over 
{M_{2}^{2}}},\nonumber\\
l'^{2} & = & 
{{\left(v^{2} - 4 f^{2} m_{l}^{2} \right ) M_{2}^{2}} \over {8 f^{4} \left (
M_{2}^{2} - M^{2} \right )}},\ 
V = - f^{2} \left ( 1 - {{M^{2}} \over{M_{2}^{2}}} \right )l'^{4}.
\eea

If we assume that only $M'\approx 0$, we find a parity breaking solution
(given that $l\neq 0,\ l'\neq 0$): 
\bea{mysol1}
\delta & = & -{l_1^2v\over 4f^2l_1^2+M_1^2},\  
\delta'=-{l_2^2v\over 4f^2l_2^2+M_1^2}\nonumber\\
\viiva{\delta} & = & -{l_1^2M\over M_2^2},\ 
\viiva{\delta}'=-{l_2^2M\over M_2^2}
\eea
and $l^2, l'^2$ can be solved from equations
\bea{mysol2}
m_l^2-{v^2\over 2f^2}(1-{M_1^2\over 4f^2l^2})+2f^2(1-{M^2\over M_2^2}+{v^2\over 8f^4}(1-{M_1^2\over 2f^2l^2})l^2+
h^2l'+\cO(M_1^4) & = & 0\nonumber\\
m_l^2-{v^2\over 2f^2}(1-{M_1^2\over 4f^2l'^2})+2f^2(1-{M^2\over M_2^2}+{v^2\over 8f^4}(1-{M_1^2\over 2f^2l'^2})l'^2+
h^2l+\cO(M_1^4) & = & 0.\nonumber\\
\eea
We find that in this case $l$ generally does not vanish. 
Numerical examinations, when further requiring that $M'\neq 0$, support this result.

Note that after parity breaking the aqcuired VEVs may change as
temperature falls and electroweak symmetry is broken by the VEVs 
of the
$\Phi$ and $\chi$ fields. The $\Phi$ and $\chi$ fields acquire $Q_{em}$
conserving VEVs, \be{phivev} \Phi=\left(\matrix{ \kappa_1 & 0 \cr 0 &
\kappa_1' \\ }\right),\ \chi=\left(\matrix{ \kappa_2' & 0 \cr 0 & \kappa_2
\\ }\right) \ee where $\sqrt{\kappa_1^2+\kappa_2^2}$ is of the order of
$\sim 100$ GeV after the EW phase transition and $\kappa_i'\ll\kappa_i$. 
$\kappa_1$ and $\kappa_2$ can be chosen real using $U(1)$-invariance. 

To study the behaviour of the sneutrino VEV at the phase transition, we set
the fields to their VEVs after the EW breaking (we also approximate $\delta=0,\
\viiva{\delta}=0$ and set $\kappa_1'=\kappa_2'=0$ for simplicity). The
potential (\ref{higpot}) simplifies to
\bea{ewpot} 
V & = & \abs{h_\chi
l\kappa_2+2fl'\delta'}^2+\abs{h_\chi l'\kappa_2}^2+ \abs{h_\phi
ll'+2\mu_1'\kappa_1}^2+\abs{\mu_1\kappa_2}^2 +|h_\chi
l'l+\mu_1\kappa_1|^2\nonumber\\ & + &
4|\mu_1''\kappa_2|^2+|f|^2|l|^4+|fl'^2+M\viiva{\delta}'|^2+
m_l^2(|l|^2+|l'|^2)+M_1^2|\delta'|^2\nonumber\\ & + &
(M_2-|M|^2)|\viiva{\delta}'|^2 + M'^2(\delta'\viiva{\delta}'+ \hc)
+(M_\phi^2-4|\mu_1'|^2)|\kappa_1|^2+
(M_\chi^2-4|\mu_1''|^2)|\kappa_2|^2\nonumber\\ & + & ({\mu_2^2\over
2}\kappa_1\kappa_2+\hc)+(vl'^2\delta'+e_2l\kappa_2 l'+\hc) +{g_L^2\over
8}\Big\arrowvert|l|^2+|\kappa_1|^2-|\kappa_2|^2\Big\arrowvert^2\nonumber\\
& + & {g_R^2\over 8}\Big\arrowvert|l'|^2+2(|\viiva{\delta}'|^2-|\delta'|^2
+|\kappa_1|^2-|\kappa_2|^2)\Big\arrowvert^2
\nonumber\\
& + &  {g'^2\over 8}\Big\arrowvert|l'|^2-|l|^2+2(|\viiva{\delta}'|^2-|\delta'|^2)\Big\arrowvert^2.
\eea
Furthermore, assuming that $f,M,v,e_2,\delta',\viiva{\delta}'$ are real (for simplicity)
and setting $h_\phi=0$ the terms containing $l'$ are
\bea{ewpot2}
V_{l'} & = & 4f^2|l'|^2\delta'^2+4fl\kappa_2\delta'\Re(h_\chi l')
+\kappa^2|h_\chi|^2|l'|^2+|h_\chi|^2|l'|^2|l|^2+
2\kappa_1l\mu_1\Re(h_\chi l')\nonumber\\
& + & f^2|l'|^4+2fM\viiva{\delta}'\Re(l'^2)
+ m_l^2|l'|^2+2v\delta'\Re(l'^2)+2e_2l\kappa_2\Re(l')\nonumber\\
& + & {g_R^2\over 8}[|l'|^4+2|l'|^2(2\viiva{\delta}'^2-2\delta'^2+\kappa_1^2-
\kappa_2^2)]\nonumber
\\
& + & {g'^2\over 8}[|l'|^4+2|l'|^2(2\viiva{\delta}'^2-2\delta'^2)-2|l'|^2|l|^2].
\eea
Writing $l'=|l'|e^{i\alpha},h_\chi=h e^{i\beta}$ and 
differentiating with respect to the
phase of the complex VEV, $\alpha$, we obtain
\bea{myder}
-\dd{V_{l'}}{\alpha} & = & 
[4hfl\kappa_2\delta'|l'|+2h\mu_1l|l'|\kappa_1]\sin(\alpha+\beta)+
4(fM\viiva{\delta}'+v\delta')|l'|^2\sin(2\alpha)\nonumber\\
& + & 2e_2l\kappa_2|l'|\sin(\alpha)
\eea
($|l'|$ can be assumed to be approximately constant to ensure parity breaking
and the required heavy right-handed particle masses).
We note that for $\alpha\neq 0$ we must require that $l\neq 0$.
If $\beta=0$ we may easily solve $\dd{V_{l'}}{\alpha}=0$ for $\alpha$:
\be{alpha}
\alpha_1=\arccos(-{2hfl\kappa_2\delta'+h\mu_1l\kappa_1+e_2l\kappa_2\over
4(fM\viiva{\delta}'+v\delta')|l'|}).
\ee

To examine when this is a minimum we substitute $\alpha_1$ into 
(\ref{ewpot2}) (writing only the $\alpha$-dependent terms):
\be{apot1}
V_{l'}(\alpha=\alpha_1)=-
{(2hfl\kappa_2\delta'+h\mu_1l\kappa_1+e_2l\kappa_2)^2\over 4(
fM\viiva{\delta}'+v\delta')}-2fM\viiva{\delta}'l'^2-2v\delta'l'^2.
\ee
At $\alpha=0$ (the old minimum) (\ref{ewpot2}) is (again writing only the 
$\alpha$ dependent part):
\bea{apot2}
V_{l'}(\alpha=0) & = & 4fl\kappa_2\delta'h l'+2\kappa_1l\mu_1h l'\nonumber\\
& + & 2fM\viiva{\delta}'l'^2
+ 2v\delta'l'^2+2e_2l\kappa_2l'.
\eea
To assure that $\alpha_1$ is a minimum, we must thus require that
\be{cond1}
V_{l'}(\alpha_1)<V_{l'}(0)
\ee
from which it follows that
\bea{cond2}
4hfl\kappa_2\delta'l'+2h\mu_1l\kappa_1l'+2e_2l\kappa_2l'+
4fM\viiva{\delta}'l'^2+4v\delta'l'^2
& + &\\
{(2hfl\kappa_2\delta'+h\mu_1l\kappa_1+e_2l\kappa_2)^2\over 4(fM\viiva{\delta}'
+v\delta')} & < & 0.
\eea
If $\beta\neq 0$ i.e. $h_\chi$ is complex, this question cannot be answered
analytically and numerical methods must be utilized.

Provided that condition (\ref{cond2}) is satisfied, it is then clear that 
as $\Phi$ and $\chi$ acquire their VEVs at the electroweak 
transition the sneutrino VEV, $l'$,
while remaiming approximately constant in magnitude rotates 
in the complex plane by an angle $\alpha_1$. 

Expanding around the new minimum, we may write
\be{lvev1}
l'_{EW}\equiv l'e^{i\alpha}=l'+Ae^{i\theta},
\ee
from which it follows that 
\be{lvev2}
l'\sqrt{2(1-\cos\alpha)}
e^{i\arctan({\sin\alpha\over\cos\alpha-1})}=Ae^{i\theta}.
\ee
In the EW phase transition, $A$ changes 
from $0$ to $l'\sqrt{2(1-\cos\alpha)}$ while the change in $\theta$ is
$\Delta\theta=\arccot({\sin\alpha\over 1-\cos\alpha})$.
We estimate the changes in $A$ and $\theta$ to be sinusoidal
and choose that only one of the sneutrino VEVs is significantly large.
In \cite{multa2} it was shown
that for $T\approx\mu\sim 100$ GeV $\cI^{e_j}_{\mato{H}}$ is approximately
$-2$ regardless of the considered generation so that (\ref{int1}) can be
written as 
\be{i1}
I_1\approx 48 h^2l'^2(1-\cos\alpha)\arccot({\sin\alpha\over 1-
\cos\alpha})\int_0^\infty du g^2(u)g'(u)e^{-\lambda_+u},
\ee
where 
\be{gu}
g(u)={1\over 2}[1-\cos({u\pi\over L})][\theta(u)-\theta(u-L)]+\theta(u-L)
\ee
and $h\sim h_{3i},\ i=1,2,3$. 

For $T=100$ GeV, $L=25/T$ the integral has a value of $\sim 0.26$.
 To estimate the value of $I_2$ let us assume further
that $l'\sim 1$ TeV and $h\sim 0.1$ so that
\be{ni1}
I_1\approx (1-\cos\alpha)\arccot({\sin\alpha\over 1-\cos\alpha})
\ 1.2\times 10^5.
\ee 

The effect of adding a mass insertion can easily be calculated.
Taking $\abs{l'}$ to be constant, (\ref{int1})
reduces to 
\be{int2} -2 h_{ij}h_{kj}\int_0^\infty du
\cos(\theta_i-\theta_k)A_iA_k(\theta_i+\theta_k)'
e^{-\lambda_+u}\cI^{e_j}_{\mato{H}}. 
\ee 
If only one of the sneutrino fields, $A_3e^{i\theta_3}$,
has a significant VEV $\sim 1$ TeV,
(\ref{int2}) becomes

\be{int3} 
I_2\approx 24 h^2l'^2\int_0^\infty du \theta_3' e^{-\lambda_+u},
\ee 
where $l'$ is the magnitude of the sneutrino VEV and
$\cI^{e_j}_{\mato{H}}\approx -2$ as before. 

The phase of
the sneutrino field VEV is again assumed to change sinusoidally during 
the EW phase transition i.e. 
\be{sinusoid} 
\theta_3(u)=\alpha g(u)
\ee 
so that the integral in
(\ref{int3}) can be evaluated and is found to be 
\be{intvalue} 
{\pi^2\over
2(\lambda_+^2L^2+\pi^2)}(1+e^{-\lambda_+L})\alpha
\ee 
For $L=25/T$, $T=100$ GeV and 
$\lambda_+\approx 1.5$ this has a value of about $0.83\ \alpha$. Taking
$h\sim 0.1$ and $l'\sim 1$ TeV, we arrive at an estimate for
CP-violation during the electroweak phase transition with one mass insertion, 
\be{final}
I_2\approx 2\times10^5\ \alpha. 
\ee 
Clearly, since
$$(1-\cos\alpha)\arccot({\sin\alpha\over 1-\cos\alpha})\leq\alpha$$
for $\alpha\in[0,\pi]$, $I_1<I_2$
indicating that further mass insertions may possibly increase the amount of CP-violation
even further.

In comparison with $I_2$ the estimate for the
CP-violation in the MSSM \cite{CQRVW, multa2} is $\sim 300$ (without the
suppression factor \cite{multa1}) that leads to a baryon to entropy ratio
of ${n_B\over s}\sim 10^{-11}$. Because of (\ref{nbovers}) the baryon to 
entropy ratio is then 
\be{finalnbs} 
{n_B\over s}\sim \alpha\ 0.7\times 10^{-8}
\ee 
so that even a change of $\cO(10^{-3})$ in the phase of the sneutrino
VEV may lead to significant baryon production at the EW phase transition. 
This result is obtained after quite a number of assumptions and
simplifications. However, these are not carefully chosen to obtain
a desired result but are used to simplify the calculation process.

In this paper we have considered the possibility of electroweak baryogenesis 
in the supersymmetric left-right model. As in \cite{multa2} the
crucial property
that allows for sufficient baryon creation is the large sneutrino VEV.
In SUSYLR this is set quite naturally by the left-right parity
breaking scale. As electroweak breaking takes place the sneutrino
VEV must necessarily change for baryogenesis to be possible.
We have shown that for a zero temperature approximation
there exists a part of the parameter space
where the sneutrino VEV rotates in the complex plane while
conserving its large amplitude. This process creates a CP-violating
source that leads to a production of baryons at the electroweak
phase transition through a known mechanism \cite{CQRVW}.
The amount of baryons produced in this manner is
obviously dependent on many unknown parameters 
and may change significantly due to finite temperature corrections 
but even a change of $\cO(10^{-3})$ in the phase of the neutrino VEV may 
lead to a significant production of baryons.\\
\\
\noindent{\bf Acknowledgement.}\\
\noindent
The authors thank R. Mohapatra for dicussions.
\newpage
%%%%%%%%%%%%%%%%%%%%%%%%%%%%%%%%%%%%%%%%%%%%%%%%%%%%%%%%%%%%%%%%%%%%%%%% 

\end{document}